\documentstyle[12pt,epsf]{article}

\def\thefootnote{\fnsymbol{footnote}}

\newcommand{\eq}{\begin{equation}}
\newcommand{\en}{\end{equation}}
\newcommand{\eqa}{\begin{eqnarray}}
\newcommand{\ena}{\end{eqnarray}}

\newcommand{\br}{\langle}
\newcommand{\kt}{\rangle}

\def\spose#1{\hbox to 0pt{#1\hss}}
\def\ltapprox{\mathrel{\spose{\lower 3pt\hbox{$\mathchar"218$}}
 \raise 2.0pt\hbox{$\mathchar"13C$}}}
\def\gtapprox{\mathrel{\spose{\lower 3pt\hbox{$\mathchar"218$}}
 \raise 2.0pt\hbox{$\mathchar"13E$}}}

\newcommand{\JP}[1]{J.\ Phys.\ {\bf #1}}

\newcommand{\PR}[1]{Phys.\ Rev.\ {\bf #1}}

\begin{document}
\begin{titlepage}
\vskip0.5cm
\begin{flushright}
DFTT 12/2000\\
HU-EP-00/12\\
IFUP-TH 7/2000\\
Roma1-1289/00\\
\end{flushright}
\vskip0.5cm
\begin{center}
{\Large\bf High-precision estimate of $g_4$}
\vskip 0.3cm
{\Large\bf in the 2D Ising model}
\end{center}
\vskip 1.3cm
\centerline{
Michele Caselle$^a$, Martin Hasenbusch$^b$, Andrea Pelissetto$^c$ and 
Ettore Vicari$^d$}

 \vskip 0.4cm
 \centerline{\sl  $^a$ Dipartimento di Fisica Teorica dell'Universit\`a di
 Torino and I.N.F.N., I-10125 Torino, Italy}
 \centerline{\sl $^b$ Humboldt Universit\"at zu Berlin, Institut f\"ur Physik,
 Invalidenstr. 110, D-10115 Berlin, Germany}
 \centerline{\sl  $^c$ Dipartimento di Fisica dell'Universit\`a di Roma I
 and I.N.F.N., I-00185 Roma, Italy}
 \centerline{\sl  $^d$ Dipartimento di Fisica dell'Universit\`a di Pisa 
 and I.N.F.N., I-56127 Pisa, Italy}
 \vskip 1.cm

\begin{abstract}
We compute the renormalized four-point coupling in the 2d Ising model
using transfer-matrix techniques. We greatly reduce the systematic 
uncertainties which usually affect this type of  calculations by using
the exact knowledge of several terms in the scaling function of the 
free energy. Our final result is $g_4=14.69735(3)$.

\end{abstract}
\end{titlepage}

\setcounter{footnote}{0}
\def\thefootnote{\arabic{footnote}}

\section{Introduction}
An important information on the physical properties of a quantum field 
theory is
given by the renormalized four-point coupling, which is defined in terms of 
the zero momentum projection of the truncated 4-point correlator. 
At the same time, if one is interested in the lattice discretization of 
the theory, this renormalized coupling represents one of the
most interesting universal amplitude ratios, being related to the fourth
derivative of the free energy.

Recently, in~\cite{b2000}, a new interesting  approach 
has been proposed to evaluate this quantity in the case of integrable QFT's.
The idea is that for these theories one has direct access 
to the so called form factors from which the renormalized coupling 
can be computed.  In~\cite{b2000} the method was tested in the case of the 
2d Ising model.  The authors found the remarkably precise estimate
\eq
g_4^*=14.6975(1)
\en
(see below for the precise definition of $g_4^*$).

The aim of this paper is to test this result by using a completely different
method. By combining transfer-matrix methods and the exact knowledge of 
several
terms in the scaling function of the free energy  of the model we are able to
obtain a precision similar to that of~\cite{b2000}. Our result is
\eq
g_4^*=14.69735(3),
\en
which is in substantial agreement with the estimate of~\cite{b2000}.
Given the subtlety of the calculations involved in both 
approaches, our result represents a highly non-trivial test of both
methods. In performing our analysis we employ the same techniques which were 
used in~\cite{ch99} in the study of the 2d Ising model in a magnetic field. 

This paper is organized as follows:
We begin in sect.2 by collecting some definitions and elementary results
which will be useful in the following. 
Sect.3 is devoted to a discussion of the
transfer-matrix results (and of the techniques that we use to improve the
performance of the method). In sect.4 we obtain the first 4 terms
of the scaling function for the fourth derivative of the free energy, which
enters in the estimate of $g_4$ and finally, in sect.5, we discuss the fitting
procedure that we used to extract the continuum-limit value from the data. 
To help the reader to follow our analysis, we have listed in tab.\ref{tab1} 
the output of our transfer-matrix analysis.

\section{General setting}

We are interested in the 2d Ising model defined
by the partition function
\eq
Z=\sum_{\sigma_i=\pm1}e^{\beta\sum_{\br n,m \kt}\sigma_n\sigma_m
+h\sum_n\sigma_n} ,
\label{zz1}
\en
where the field variable $\sigma_n$ takes the values $\{\pm 1\}$;
$n\equiv(n_0,n_1)$ labels the sites of a square lattice of size $L_0$ and $L_1$
in the two directions
and $\br n,m \kt$ 
denotes nearest neighbour sites on the lattice.
In our calculations with the transfer-matrix
method we shall treat asymmetrically the two directions. We shall denote
$n_0$ as the ``time'' coordinate and $n_1$ as the ``space" one.
The number of sites  of
the lattice will be denoted by  $N\equiv L_0 L_1$.
The critical value of $\beta$ is
 $$\beta=\beta_c=\frac12\log{(\sqrt{2}+1)}=0.4406868...$$
In the following we shall be interested in the high-temperature phase of the
model in which the ${\bf Z_2}$ symmetry is unbroken, i.e. in the region
$\beta<\beta_c$. It is useful to introduce the reduced temperature $t$
 defined as:
\eq
t\equiv \frac{\beta_c-\beta}{\beta_c} .
\en
As usual, we introduce the free-energy density $F(t,h)$ and 
the magnetization per site $M(t,h)$ defined as
\eq
F(t,h)\equiv {1\over N} \log(Z(t,h)),    \qquad\qquad
M(t,h)\equiv {\partial F(t,h)\over \partial h}.
\en
The standard definition of the four-point zero-momentum
renormalized coupling constant $g_4$ is
\eq
g_4(t)=-\frac{F^{(4)}}{\chi^2\xi_{\rm 2nd}^2},
\en
where
$\chi$ and $F^{(4)}$ are the second- and fourth-order derivatives of
the free-energy density $F(h,t)$ at $h=0$:
\eq
\chi(t)=\left. \frac{\partial^2 F(t,h)}{(\partial h)^2}\right\vert_{h=0},
\qquad\qquad
F^{(4)}(t)=\left. \frac{\partial^4 F(t,h)}{(\partial h)^4}\right\vert_{h=0}\; 
\en
and  $\xi_{\rm 2nd}$ denote the second moment correlation length. The 
second moment correlation length is defined by
\eq
\xi_{\rm 2nd}^2 = \frac{\mu_2}{2 d \mu_0} \;,
\en
where $d$ is the dimension (here $d=2$) and
\begin{equation}
\mu_i = \lim_{L_1 \rightarrow \infty} \lim_{L_0 \rightarrow \infty}
      \frac{1}{N} \sum_{m,n} (m-n)^i <\sigma_m \sigma_n>_c .
\end{equation}
The connected part of the correlation function is given by
\begin{equation}
 <\sigma_m \sigma_n>_c = <\sigma_m \sigma_n>   - <\sigma_m> <\sigma_n> .
\end{equation}

In particular, we are interested in the continuum-limit value $g_4^*$
defined as
\eq
g_4^*=\lim_{t\to0}g_4(t).
\en
For $t\to 0$ we have
\eq
\xi_{\rm 2nd}(t) \simeq A_{\xi,\rm 2nd}\ t^{-1}, \qquad\qquad
\chi (t)\simeq A_{\chi}\ t^{-7/4}, \qquad\qquad
F^{(4)}(t) \simeq A_{F^{(4)}}\ t^{-11/2}, 
\en
from which it follows
\eq
g_4^*=-\frac{A_{F^{(4)}}}{A_{\chi}^2A_{\xi,\rm 2nd}^2}.
\label{g4last}
\en
The amplitude $A_\chi$ is known exactly (see e.g. ref.~\cite{mccoy}): 
$A_\chi = 0.9625817322...$ The amplitude $A_{\xi,\rm 2nd}$ can also be 
computed exactly. Indeed, consider the exponential correlation 
length $\xi$ (inverse mass gap). For $t\to 0$, it behaves as \cite{mccoy} 
$A_\xi t^{-1}$, with $A_\xi = 1/(4 \beta_c) = 0.56729632855....$
Using then \cite{CPRV} $A_\xi/A_{\xi,\rm 2nd} = 1.000402074...$, 
we obtain finally $A_{\xi,\rm 2nd}= 0.5670683251....$ 

Our goal in the remaining part of this paper is to give a numerical estimate 
of $A_{F^{(4)}}$.

\section{Transfer-matrix results}

We may have direct numerical access to
${F^{(4)}}$ by looking at the $h$ dependence of the magnetization at fixed $t$.
Expand as follows:
\eq
h=b_1M+b_3 M^3 \cdots
\label{1.1}
\en
 we immediately see that
\eq
b_1=1/\chi, \qquad\qquad
b_3= -\frac{F^{(4)}}{6 \chi^4},
\en
so that
\eq
F^{(4)}=-6b_3/b_1^4.
\en

\subsection{The transfer-matrix technique}

As a first step we computed the magnetization $M$ of a system with 
$L_0=\infty$ and finite $L_1$. 
The magnetization of this system is given by
\begin{equation}
 M =  v_0^T \tilde M v_0 \;\;,
\end{equation}
where $v_0$ is the eigenvector of the transfer matrix % \cite{KrWa}
with the largest eigenvalue and 
$\tilde M$ is a diagonal matrix with $\tilde M_{ii}$ being equal 
to the magnetization of the time-slice configuration $i$. 
For a detailed discussion of the transfer-matrix method 
see e.g. ref. \cite{CaFi}.

We computed $v_0$ using the most trivial iterative method,
\begin{equation}
v^{n+1}_0 = \frac{T v^{n}_0}{|T v^{n}_0|},
\end{equation}
starting from a vector with all entries being equal.

An important ingredient in the calculation is the fact that the transfer
matrix can be written as the product of sparse matrices
(see e.g. ref. \cite{Ni90}).
This allows us to reach $L_1=24$ on a workstation. The major limitation  is
the memory requirement. We have to store two vectors of size $2^{L_1}$.
Since we performed our calculation in double precision, this means that
268 MB are needed. Slightly larger $L_1$ could be reached by using a
super-computer with larger memory space.

For the parameters $\beta$ and $h$ that we studied,  $n \le 200$ was
sufficient to converge within double-precision accuracy.

\subsection{The equation of state}

In order to obtain high-precision estimates of $F^{(4)}$ it turns out to be 
important to consider the external field $h$ as a function of the magnetization 
rather than the opposite.
The advantage of the series~(\ref{1.1}) is that the coefficients 
--- at least those we can compute --- are all positive,
and therefore, truncation errors are less severe than in the case of $m(h)$.

There is no sharp optimum in the truncation order.
After a few numerical experiments we decided to keep in 
eq. (\ref{1.1}) the terms up to $b_{15} \; M^{15}$:
\begin{equation}
\label{trunc}
 h(M) \;=\; b_1 \; M \;+\; b_3 \; M^3 \; + \; \cdots  \;+\; b_{15} M^{15} 
\;\;\;.
\end{equation}
In order to compute the coefficients $b_1$, $b_3$, ..., $b_{15}$ we 
solved the system of linear
equations that results from inserting 8 numerically calculated values of the 
magnetization $M(h_1)$, $M(h_2)$, ..., $M(h_8)$ into the truncated 
equation of state~(\ref{trunc}). Here we have chosen $h_j = j \; h_1$.

The errors introduced by the truncation of the series decrease as $h_1$ 
decreases, while the errors from numerical rounding 
increase as
$h_1$ decreases. Therefore, we varied $h_1$ to find the optimal choice.
For a given value of $\beta$ we performed this search only for one lattice 
size $L_1$. (Typically $L_1=18$). From the variation of the result with $h_1$ 
we can infer the precision of our estimates of $b_i$.
For example, for $\beta=0.37$,  we get $b_1$ with 14
significant digits and  $b_3$ with 12 significant digits.

\subsection{Extrapolation to the thermodynamic limit}

\begin{table}
\caption{\sl \label{iterative}
Extrapolation of $b_3$ to the thermodynamic limit for $\beta=0.37$.
Iterative procedure. The numbers in the top row give the 
extrapolation level. For the discussion see the text. }
\begin{center}
\begin{tabular}{|c|c|c|c|c|c|}
\hline
$L_1$  & 0  &   1     &    2     &    3     &             4  \\
\hline
13 & 0.0459057204193 &             &              &              &          \\
14 & 0.0463456447150 &             &              &              &          \\
15 & 0.0467262982921 & 0.049170965 &              &              &             \\
16 & 0.0470483839889 & 0.048819648 &              &              &           \\
17 & 0.0473162239288 & 0.048638691 & 0.048446477  &              &           \\
18 & 0.0475358880370 & 0.048537476 & 0.048409004  &              &            \\
19 & 0.0477140164478 & 0.048477931 & 0.048392846  & 0.048380598  &            \\
20 & 0.0478571162217 & 0.048441711 & 0.048385463  & 0.048379249  &            \\
21 & 0.0479711755247 & 0.048419155 & 0.048381922  & 0.048378658  & 0.048378196 \\
22 & 0.0480614838402 & 0.048404863 & 0.048380148  & 0.048378366  & 0.048378082  \\
23 & 0.0481325804063 & 0.048395686 & 0.048379222  & 0.048378214  & 0.048378046 \\
24 & 0.0481882778282 & 0.048389731 & 0.048378721  & 0.048378128  & 0.048378019 \\
\hline
\end{tabular}
\end{center}
\end{table}

 ~From the transfer matrix formalism it follows that for periodic boundary 
conditions and $\beta\ne \beta_c$, the free energy density approaches its 
thermodynamic limit value exponentially in $L_1$. Hence, also derivatives of
the free energy density with respect to $h$ and linear combinations of them   
should converge exponentially in $L_1$ to their thermodynamic limit value.
Therefore, in the simplest case, one would 
extrapolate with an Ansatz
\eq
\label{simpleex}
 b(L_1) = b(\infty) + c \exp(-x L_1) \;\;\;,
\en
where $b(L_1)$ is the quantity at the given lattice size $L_1$ and $b(\infty)$
the thermodynamic limit of the quantity. In order to obtain numerical estimates
for $b(\infty)$, $c$ and $x$ we have inserted the numerical
result of $b$ for the three lattice sizes $L_1$, $L_1-1$ and $L_1-2$ into 
eq. \ref{simpleex}. It turns out that using this simple extrapolation, still
a dependence of the result for $b(\infty)$ on $L_1$ is visible. This indicates that,
with our numerical precision, subleading exponential corrections have to be 
taken into account. For this purpose we have iterated the extrapolation discussed
above. 

The iteration starts with  $b^{(0)}(L_1)$  which are the quantities $b$ 
that have been computed by the transfer matrix for the lattice size $L_1$.
A step of the iteration is given by solving the system of equations 
\begin{eqnarray}
b^{(i)}(L_1-2) &=& c \; \exp(-x (L_1-2)) \;+\; b^{(i+1)}(L_1)  \nonumber \\
b^{(i)}(L_1-1) &=& c \; \exp(-x (L_1-1)) \;+\; b^{(i+1)}(L_1)    \nonumber \\
b^{(i)}(L_1) \phantom{-0)}  &=& c 
\; \exp(-x L_1) \phantom{(-2x)} \;+\; b^{(i+1)}(L_1) \;\;\;.
\end{eqnarray}
with respect to  $b^{(i+1)}(L_1)$, $c$ and $x$.
In table \ref{iterative} we give as an example
the extrapolation of $b_3$ at $\beta=0.37$.
In the second column we give the results obtained for the given $L_1$.
The stability of the extrapolation with varying $L_1$ increases up to the 
fourth iteration. Further iterations become numerically unstable.

As final result we took $b_3=0.04837802(3)$ from the $4^{th}$ iteration.
The error was estimated from the variation of the results with $L_1$. As  
a consistency check, we also extracted the thermodynamic limit by fitting 
with multi-exponential Ans\"atze. We found consistent results.  The 
relative accuracy of $b_1$ in the 
thermodynamic limit was in general better than that of $b_3$.

In the second column of table \ref{tab1} we give 
our final results for $-F^{(4)} t^{11/2}$ at the $\beta$ values that we 
have studied.
For a discussion of the following columns 
see section 5.

\begin{table}
\caption{\sl \label{tab1}
Extrapolation of $a_{F_4}$ }
\begin{center}
\begin{tabular}{|c|l|l|l|l|l|}
\hline
$\beta$ &$-F^{(4)} t^{11/2}$  & $\frac{-F^{(4)} u_t^{11/2}} {(u_h/h)^{4}}$  &
$+b_{F_4} t^{2.75} $
& ext.  $t^{3.75}$  &  ext.  $t^4$  \\
\hline
0.200 &3.21111498292(2) &  4.202506 & 4.358692  &            &           \\
0.250 &3.4721394466(1)  &  4.286829 & 4.369154  &  4.383609  & 4.382846 \\
0.280 &3.6225400346(6)  &  4.321965 & 4.373381  &  4.381365  & 4.380954  \\
0.300 &3.720514859(3)   &  4.339751 & 4.375424  &  4.380395  & 4.380141 \\
0.310 &3.768883189(7)   &  4.347109 & 4.376235  &  4.380086  & 4.379911 \\
0.320 &3.81687386(2)    &  4.353517 & 4.376917  &  4.379774  & 4.379635 \\
0.330 &3.86451569(5)    &  4.359033 & 4.377479  &  4.379545  & 4.379438 \\
0.340 &3.91183946(6)    &  4.363716 & 4.377934  &  4.379382  & 4.379302 \\
0.350 &3.9588780(3)     &  4.367629 & 4.378292  &  4.379270  & 4.379213  \\
0.355 &3.9823015(7)     &  4.369315 & 4.378439  &  4.379239  & 4.379196 \\
0.360 &4.0056667(8)     &  4.370833 & 4.378566  &  4.379203  & 4.379168 \\
0.365 &4.028978(1)      &  4.372189 & 4.378674  &  4.379174  & 4.379146 \\
0.370 &4.052240(2)      &  4.373391 & 4.378765  &  4.379148  & 4.379126 \\
\hline
\end{tabular}
\end{center}
\end{table}

\section{Scaling function for $F^{(4)}$}

In this Section we shall study the asymptotic behavior 
of  $F^{(4)}(t)$ for $t\to 0$
following Ref.~\cite{af}. With respect
to~\cite{af}, we have added the contributions due to the
irrelevant operators. Here, we shall use the knowledge of the operator
content of the theory at the critical point which can be obtained by using the
methods of 2d conformal field theories.

General renormalization-group (RG) 
arguments indicate that the free energy of the 
model can be written as 
\begin{eqnarray}
F(t,h) & = & F_{b}(t,h) + 
   |u_t|^{d/y_t} f_{\rm sing}\left(\frac{u_h}{|u_t|^{y_h/y_t}},
       \left\{\frac{u_j}{|u_t|^{y_j/y_t}} \right\}\right) 
\nonumber \\
&& +
   |u_t|^{d/y_t} \log |u_t| 
      \widetilde{f}_{\rm sing}\left(\frac{u_h}{|u_t|^{y_h/y_t}},
       \left\{\frac{u_j}{|u_t|^{y_j/y_t}} \right\}\right).
\end{eqnarray}
Here $F_{b}(t,h)$ is a regular function of $t$ and $h^2$,
the so-called bulk contribution, $u_t$, $u_h$, $\{u_j\}$ are the 
non-linear scaling fields associated respectively to the 
temperature, the magnetic field and the irrelevant operators, and 
$y_t$, $y_h$, $\{y_j\}$ are the corresponding dimensions. 
For the Ising model $y_t = 1$, $y_h= 15/8$. Notice the 
presence of the logarithmic term, that is related to a ``resonance"
between the thermal and the identity operator\footnote{In principle, 
logarithmic terms may also arise from additional 
resonances due to the fact 
that $y_j$ are integers or differ by integers from $y_h$. They will not 
be considered here since these contributions either are subleading with 
respect to those we are interested in or have a form that is already included.}.
The scaling fields are analytic functions of $t$ and $h$ that 
respect the ${\bf Z_2}$ parity of $t$ and $h$. Let us write the Taylor expansion
for $u_h$ and $u_t$, keeping only those terms
that are needed for our analysis (we use the notations of~\cite{af}):
\eq
u_h~=~h~[1~+~c_ht~+~d_ht^2~+~e_h h^2~+~ f_h t^3 ~+~O(t^4,th^2)],
\label{u_h}
\en
\eq
u_t~=~ t ~+~b_t h^2 ~+~c_tt^2 ~+~d_t t^3 ~+~e_tth^2~+~ g_t h^4 ~+~
~ f_t t^4 ~+ O(t^5,t^2h^2).
\label{u_t}
\en
Let us first discuss the contributions of the irrelevant operators.
In generic models their dimensions are usually unknown. In the present case
instead, we may identify the irrelevant operators with the secondary fields
obtained from the exact solution of the model at the critical point
and  use the corresponding RG exponents as
input of our analysis. We shall discuss this issue in full detail in a
forthcoming publication, let us only summarize here the main results of this
analysis. It turns out that, discarding corrections of order $O(t^5)$,
we have only two possible contributions:
\begin{itemize}
\item
The first one is due to terms  $T\bar T$ , $T^2$ and $\bar T^2$
(where $T$ denotes the energy-momentum tensor). These terms
would give a correction proportional to $t^2$ in the scaling function.
\item
The second contribution is due to the $L_{-3}\bar
L_{-3} I$ field from the Identity family and to $L_{-4}\epsilon$,
$\bar L_{-4}\epsilon$ from the energy family (where the $L_{-i}$'s are
the generators of the Virasoro algebra).
These terms give a correction proportional to $t^4$ in the scaling function.
\end{itemize}
However, it turns out (see for instance the remarks of~\cite{ch99,af,Nickel})
that in the infinite-volume free energy of the 2d Ising model 
the $T\bar T$ , $T^2$ and $\bar T^2$
terms are actually absent\footnote{This conjecture is verified by the free 
energy and by the susceptibility at $h=0$ \cite{Nickel} and by the 
free energy $F(0,h)$ \cite{ch99}. Note that this is expected to be true 
only in the thermodynamic limit. In the finite-size scaling limit 
%corrections with $\omega=2$ are indeed observed \cite{deQueiroz}.
corrections that vanish like $L_1^{-2}$ are indeed observed \cite{deQueiroz}.
It is also not true for other observables, for instance, for the
correlation length $\xi$.}. 
Thus, from the above analysis we see that
the first correction due to the irrelevant fields appears only
at order $t^4$. Therefore, since $u_j/|u_t|^{y_j/y_t}$ vanishes for 
$t\to 0$, we can expand
\eq
f_{\rm sing} (x,\{z_j\}) = Y_+ (x) + u_0(t,h) u_t^4 X_+(x) + O(u_t^5),
\en
where $u_0(t,h)$ is an analytic function of $t$ and $h$, and $Y_+$, 
$X_+$ are appropriate scaling functions. The same expansion holds 
for $\widetilde{f}_{\rm sing}$ with different functions
$\widetilde{Y}_+$, $\widetilde{X}_+$. 
Additional constraints can be obtained using the exactly known results 
for the free energy, the magnetization and the susceptibility in zero field. 
Since all
numerical data indicate that all zero-momentum correlation functions 
diverge as a power of $t$ without logarithms for $t\to 0$, we assume as
in Ref. \cite{af} 
that $\widetilde{Y}_+(x)$ is constant, i.e. 
$\widetilde{Y}_+(x) = \widetilde{Y}_0$. The exact results for the 
free energy and the magnetization give then~\cite{ssv1}
\eq
c_h=\frac{\beta_c}{\sqrt{2}},  \hskip1cm
d_h=\frac{23 \beta_c^2}{16},   \hskip1cm
f_h=\frac{191 \beta_c^3}{48\sqrt{2}},
\label{eq24}
\en
\eq
c_t=\frac{\beta_c}{\sqrt{2}},\hskip1cm
d_t=\frac{7 \beta_c^2}{6}, \hskip1cm
f_t=\frac{17 \beta_c^3}{6\sqrt{2}},
\label{eq25}
\en
where we have adapted the numbers of \cite{ssv1} to our normalizations,
and $\widetilde{Y}_0 = - 4 \beta_c^2/\pi$. 
By making use of the expansion of the susceptibility, we obtain further
\eq
Y^{(2)}_+ (0) = A_\chi, \qquad \qquad b_t = - {D_0 \pi\over 16 \beta_c^2},
\en
where $D_0$ is the coefficient of the contribution proportional 
to $t\log |t|$ in the susceptibility. Numerically $D_0 = 0.04032550\ldots$,
so that $b_t = -0.0407708\ldots$ Nickel \cite{Nickel} has also 
conjectured, on the basis of the numerical analysis of the 
high-temperature series of the susceptibility, that 
$e_t = b_t \beta_c \sqrt{2}$.

Using the results presented above, and taking four derivatives of the 
free energy we obtain
\begin{eqnarray}
F^{(4)}&=& t^{-11/2}\ (a_{F4}(t)\ + t^4 \widetilde{a}_{F4}(t) \log |t|)\ 
       + t^{-11/4}\ (b_{F4}(t)\ + t^4 \widetilde{b}_{F4}(t) \log |t|)
\nonumber \\
       && + c_{F4}(t)\ + \widetilde{c}_{F4}(t) \log |t|,
\label{1.4}
\end{eqnarray}
where  $a_{F4}(t)$, $b_{F4}(t)$, $c_{F4}(t)$, 
       $\widetilde{a}_{F4}(t)$, $\widetilde{b}_{F4}(t)$,
   and $\widetilde{c}_{F4}(t)$ 
are analytic functions. Using Eqs.~(\ref{eq24}) and (\ref{eq25}),
we can compute the first terms in the Taylor expansion of $a_{F4}(t)$.
By direct evaluation we find 
\begin{eqnarray}
a_{F4}(t) & = & Y_+^{(4)}(0) \frac{(1+c_ht+d_ht^2+f_ht^3)^4}
{(1+c_tt+d_tt^2+f_tt^3)^{11/2}}+ O(t^4)   
\nonumber \\
  & = & Y_+^{(4)}(0) \left(1-\frac{3\beta_c}{2\sqrt{2}}t
+\frac{13\beta_c^2}{48} t^2
+\frac{29\beta_c^3}{32\sqrt{2}} t^3 \right)
+ O(t^4).
\label{x2}
\end{eqnarray} From Eq. (\ref{x2}), we immediately identify
\eq
Y_+^{(4)}(0)=A_{F^{(4)}}.
\en
Analogously, a direct calculation shows that 
\eq
b_{F4}(0) =\,  -21\ b_t\ Y_+^{(2)}(0) = 0.8241504....
\en
The contributions proportional to $c_{F4}(t)$ and $\widetilde{c}_{F4}(t)$
give corrections of order $t^{11/2}$ which will be neglected in the following.

Putting together the various terms, we end up with 
the following expression for the
scaling function:
\begin{eqnarray}
F^{(4)} t^{11/2}
&=& A_{F^{(4)}}(1+p_1t+p_2t^2+p_3t^3) 
\nonumber \\
&& +p_4  t^{11/4} + p_5 t^{15/4}
 +p_6  t^{4} + \widetilde{p}_6 t^4 \log |t| + p_7 t^{19/4}+ O(t^5)
\label{fit}
\end{eqnarray}
where
\begin{eqnarray}
p_1&=&-\frac{3\beta_c}{2\sqrt{2}}= -0.46741893...   \\
p_2&=&\frac{13\beta_c^2}{48} =0.052597147...   \\
p_3&=&\frac{29\beta_c^3}{32\sqrt{2}} =0.054843243...   \\
p_4&=&-21b_tY_+^{(2)}(0)  = 0.8241504....   
\end{eqnarray}
and $p_5,p_6,\widetilde{p}_6,p_7$ 
and $A_{F^{(4)}}$ are undetermined constants which we shall
try to fix in the next section.

\section{Analysis of the data}
The aim of this section is to obtain a numerical estimate for  $A_{F^{(4)}}$ by
fitting the data reported in tab.\ref{tab1}
 with the scaling function (\ref{fit}).
The major problem in doing this
 is to estimate the systematic errors involved in the
truncation of the scaling function. To this end we performed two different 
types of analysis. Let us see in detail the procedure that we followed.

\subsection{First level of analysis}
We first performed a rather naive analysis of the data. In table 
\ref{tab1} we include step by step the information that we have gained in the 
previous section. In the third column of table \ref{tab1} we have multiplied
$-F^{(4)}$ by $\frac{u_t^{11/2}}{(u_h/h)^{4}}$, 
where $u_h$ and $u_t$ are given by
eqs.~(\ref{u_h},\ref{u_t}). We see that the variation from $\beta=0.30$
to $\beta=0.37$  of the numbers in column three is reduced by a factor of
 about 10 
compared with column two. In column four we add 
$b_{F4}(0) \; t^{11/4}$ to the numbers of column three. Again we see that the 
variation of the numbers with $\beta$ is drastically reduced in column four 
compared with column three.

Since we do not know the coefficients of higher order corrections exactly we
 have 
to extract them from the data. In the last two columns of table \ref{tab1} we 
have  extrapolated linearly in $t^x$, with $x=3.75$ in column 5 and $x=4$ in 
column 6. For the extrapolation we used neighboring $\beta$-values 
(e.g. the value quoted for $\beta=0.37$ is obtained from the 
extrapolation of the data for $\beta=0.365$ and $\beta=0.37$).

We see that the result of the extrapolation does not vary very much when the 
exponent is changed from $15/4$ to $4$. Also the numbers given in 
column 5 and 6 are much more stable than those of column 4.

 ~From this naive analysis we conclude that $a_{F_4}(0)=4.3791(1)$, where the 
error bar is roughly estimated from an extrapolation of column 5 with $t^4$.

In the next section we shall 
try to include the higher order corrections in a more
sophisticated fitting procedure.

\subsection{Second level of analysis}
We made three types of fits:
\begin{description}
\item{f1]}
In the first we kept $A_{F^{(4)}}$, $p_5$ and $p_6$ as free parameters.
\item{f2]}
In the second we kept $A_{F^{(4)}}$, $p_5$, $p_6$ and $p_7$ as free parameters.
\item{f3]}
In the third we kept $A_{F^{(4)}}$,
 $p_5$, $p_6$ and $\widetilde{p}_6$ 
 as free parameters.
\end{description}
These are the only choices allowed by the data. If we neglect also $p_6$ we
can never obtain an acceptable confidence level (in fact we know that $p_6$ is
certainly different from zero and our data are too precise to allow such an
approximation). If we add further terms, like a power of 
$t^5$ for instance, or try to fit simultaneously
 $p_5$, $p_6$,  $\widetilde{p}_6$ and $p_7$ it always happens that
 some of the amplitudes are  smaller than their
statistical uncertainty signalling that our data
are not precise enough to allow for five free parameters.

In order to estimate the systematic errors involved in the estimate of 
$A_{F^{(4)}}$
we performed for all the fitting functions several independent fits
trying first to fit all the existing data (those listed in tab.\ref{tab1}) 
and then eliminating the data one by one,  
starting from the farthest from the critical
point. Among the 
set of estimates of the critical amplitudes we selected only those
fulfilling the following requirements:
\begin{description}
\item{1]} \phantom{X}
The reduced $\chi^2$ of the fit must be of order unity. In order to fix
precisely a threshold we required the fit to have a confidence level larger 
than $30\%$.
\item{2]} \phantom{X}
For all the subleading terms included in the fitting function, the amplitude
estimated from the fit must be larger than the corresponding 
error, otherwise the term is eliminated from the fit.
It is exactly this constraint which forbids us to take into
account fits with more than four free parameters.
\item{3]} \phantom{X}
The amplitude of the $n^{th}$ subleading field
 must be 
such that when it is multiplied for the corresponding power of $t$, 
(for the largest value of $t$ involved in the fit) it gives a contribution
 smaller than that of the $(n-1)^{th}$ subleading term. 
This is intended to avoid artificial cancellations between
subleading fields.  
\end{description}
Among all the 
estimates of the critical amplitude $A_{F^{(4)}}$ fulfilling these requirements
we select the smallest and the largest
ones as lower and upper bounds.

The results of the fits are reported in tab.\ref{tab2}, \ref{tab3} and 
\ref{tab4}. 
We report
all the combinations of input data which fulfill requirements [1]-[3] .
In the tables we also report the best fit value of $p_5$.
All the fits were performed using the double-precision NAG routine GO2DAF.

\begin{table}[h]
\vskip 0.2cm
\begin{center}
\begin{tabular}{|l|l|c|}
\hline
\multicolumn{1}{|c|}{$A_{F^{(4)}}$}     &   
\multicolumn{1}{c|}{$p_5$}        &  d.o.f.      \\
\hline
$-4.3791092(9)$    & 0.236(3)  & 4   \\
$-4.3791065(9)$    & 0.220(12)  & 3   \\
$-4.379095(2)$    & 0.142(18) & 2   \\
\hline
\end{tabular}
\end{center}
\vskip 0.2cm
\caption{\sl Fits of type [f1] fulfilling the requirements 1-3.
 In the first column the best fit results for the critical amplitude
(with the  error induced by the systematic errors of the input
data in parenthesis),
in the second column the best fit value of $p_5$,
 in the last column
the number of degrees of freedom (i.e. the number of data used in the fit minus
the number of free parameters).}
\label{tab2}
\end{table}

\begin{table}[h]
\vskip 0.2cm
\begin{center}
\begin{tabular}{|l|l|c|}
\hline
\multicolumn{1}{|c|}{$A_{F^{(4)}}$}     &   
\multicolumn{1}{c|}{$p_5$}        &  d.o.f.      \\
\hline
$-4.3791003(1)$    & 0.0486(7)  & 7   \\
$-4.3791001(3)$    & 0.047(3)  & 6   \\
$-4.3791006(8)$    & 0.053(9)  & 5   \\
$-4.3791022(14)$   & 0.079(22)  & 4   \\
\hline
\end{tabular}
\end{center}
\vskip 0.2cm
\caption{\sl Same as tab.\ref{tab2}, but with fits of type [f2].}
\label{tab3}
\end{table}

\begin{table}[h]
\vskip 0.2cm
\begin{center}
\begin{tabular}{|l|l|c|}
\hline
\multicolumn{1}{|c|}{$A_{F^{(4)}}$}     &   
\multicolumn{1}{c|}{$p_5$}        &  d.o.f.      \\
\hline
$-4.3790944(1)$    & $-$0.692(1)  & 8   \\
$-4.3790942(3)$    & $-$0.696(7)  & 7   \\
$-4.3790961(6)$    & $-$0.634(17)  & 6   \\
$-4.3790980(10)$   & $-$0.56(4)  & 5   \\
$-4.3791005(16)$   & $-$0.41(8)  & 4   \\
\hline
\end{tabular}
\end{center}
\vskip 0.2cm
\caption{\sl Same as tab.\ref{tab2}, but with fits of type [f3].}
\label{tab4}
\end{table}

Looking at the three tables and selecting the lowest and highest values of
$A_{F^{(4)}}$ we obtain the bounds
\eq
-4.379093 \gtapprox A_{F^{(4)}} \gtapprox -4.379110,
\label{be1}
\en
from which, using eq.(\ref{g4last}), we obtain
\eq
g_4^*=14.69735(3)
\label{best}
\en
which we consider as our best estimate for $g_4^*$.
As anticipated in the introduction, this result is in substantial 
agreement with the estimate of~\cite{b2000}.
Notice however that the error quoted in eq.(\ref{best}) should not be
considered as a standard deviation. It rather encodes in a compact notation
 the systematic uncertainty of our fitting procedure.  

We can compare the estimate (\ref{best}) with previous numerical 
determinations. The analysis of high-temperature expansions gives
$g_4^*=14.694(2)$, Ref. \cite{PV-gre} and
$g_4^*=14.693(4)$, Ref. \cite{Butera-Comi}
while Monte Carlo simulations  give
$g_4^*=14.3(1.0)$, Ref. \cite{Kim-Patrascioiu}, and
$g_4^*=14.69(2)$, Ref. \cite{b2000}. 
These results agree with our estimate (\ref{best}),
which is however much more precise. 

It is clear from the data (see the second column of tab.~\ref{tab2},
\ref{tab3} and \ref{tab4}) that
the uncertainty on $A_{F^{(4)}}$ is mostly due to the
fluctuation of $p_5$. If one would be able to fix exactly also $p_5$,  
 the precision in the determination of $g_4$ could be significantly
 enhanced.

\vskip 1cm
{\bf  Acknowledgements.}
We thank Alan Sokal for useful discussions and 
Bernie Nickel for sending us his unpublished addendum, Ref.~\cite{Nickel}.
This work was partially supported by the 
European Commission TMR programme ERBFMRX-CT96-0045.

\end{document}